\documentclass[sigconf,screen, natbib=false, 9pt]{acmart}

\AtBeginDocument{%
  }

\usepackage{microtype}
\usepackage{amsmath}
\usepackage{amsfonts}
\usepackage{placeins}
\usepackage{graphicx}
\usepackage{stfloats}
\usepackage{subcaption}
\usepackage{booktabs} 
\usepackage{xspace}
\usepackage{multirow}
\usepackage{epsfig}
\usepackage{graphics}
\usepackage{times}
\usepackage{algorithm}
\usepackage{algpseudocode}

\usepackage{bm}

\usepackage{color}
\usepackage{array}
\usepackage{amsmath}
\usepackage{amsthm}
\usepackage{bm}
\usepackage{pifont}
\usepackage{comment}
\usepackage{psfrag}
\usepackage{verbatim,url}
\usepackage{float,tabularx}
\usepackage{tabularx}
\usepackage{float}

\usepackage{caption}
\usepackage{subcaption}
\usepackage{setspace}

\usepackage{pifont}
\newcommand{\cmark}{\ding{51}}%
\newcommand{\xmark}{\ding{55}}%

\usepackage{hyperref}

\algnewcommand{\LineComment}[1]{\State \(\triangleright\) #1}

\RequirePackage[
  datamodel=acmdatamodel,
  style=acmnumeric,
  ]{biblatex}

\begin{document}

\title{PARSAC: Fast, Human-quality Floorplanning for Modern SoCs with Complex Design Constraints}

\author{Hesham Mostafa}
\affiliation{%
  \institution{Intel Labs} 
  \country{USA}  
}

\author{Uday Mallappa}
\affiliation{%
  \institution{Intel Labs} 
  \country{USA}  
}

\author{Mikhail Galkin}
\affiliation{%
  \institution{Intel Labs} 
  \country{USA}  
}

\author{Mariano Phielipp}
\affiliation{%
  \institution{Intel Labs} 
  \country{USA}  
}

\author{Somdeb Majumdar}
\affiliation{%
  \institution{Intel Labs} 
  \country{USA}  
}


\begin{abstract}
  The floorplanning of Systems-on-a-Chip (SoCs) and of chip sub-systems is a crucial step in the physical design flow as it determines the optimal shapes and locations of the blocks that make up the system. Simulated Annealing (SA) has been the method of choice for tackling classical floorplanning problems where the objective is to minimize wire-length and the total placement area. The goal in industry-relevant floorplanning problems, however, is not only to minimize area and wire-length, but to do that while respecting hard placement constraints that specify the general area and/or the specific locations for the placement of some blocks. We show that simply incorporating these constraints into the SA objective function leads to sub-optimal, and often illegal, solutions. We propose the Constraints-Aware Simulated Annealing (CA-SA) method and show that it strongly outperforms vanilla SA in floorplanning problems with hard placement constraints. We developed a new floorplanning tool on top of CA-SA: PARSAC (Parallel Simulated Annealing with Constraints). PARSAC is an efficient, easy-to-use, and massively parallel floorplanner. Unlike current SA-based or learning-based floorplanning tools that cannot effectively incorporate hard placement-constraints, PARSAC can quickly construct the Pareto-optimal {\bf legal} solutions front for constrained floorplanning problems. PARSAC also outperforms traditional SA on legacy floorplanning benchmarks. PARSAC is available as an open-source repository for researchers to replicate and build on our results\footnote{\url{https://github.com/IntelLabs/parsac}}. 
\end{abstract}






\maketitle
\section{Introduction}
The design of integrated circuits (ICs) is divided into the front-end and back-end design phases. The front-end phase focuses on system specifications, architectural design choices and logic realization - in the form of gates and registers- of the circuit. The resulting logic circuit from the front-end phase is used as the starting point for the back-end design phase. The goal of the back-end phase is to provide optimal shapes and locations of logic blocks and to create interconnects. The back-end flow can be broadly divided into the following steps: partitioning, physical mapping (floorplanning or chip planning, placement, and routing), and sign-off (timing, power, and physical verification).\\

\begin{figure*}
    \centering
\includegraphics[scale=0.4]{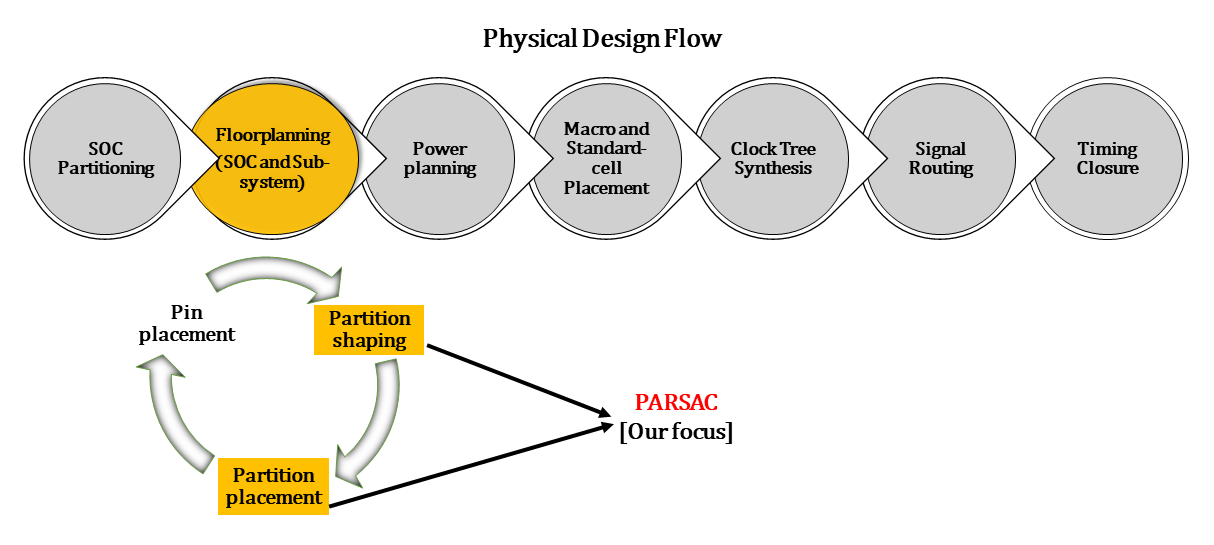}
    \caption{The SoC and sub-system floorplanning task includes hierarchy-level partition shaping, partition placement and pin placement on the partition. Our work is applicable to both SoC and sub-system level partition placement and partition shaping.}
    \label{fig:pdflow_fp}
\end{figure*}

\noindent \textbf{Motivation:} In the back-end phase, before performing physical mapping of the circuit, it is essential to partition larger flat circuit netlists into more manageable circuits or blocks. This partitioning process is predominantly influenced by factors such as the computational capabilities of Electronic Design Automation (EDA) tools, the availability of tool licenses, the size of the design team, and the time constraints for completion of the project. 
Therefore, design methodology teams use a top-down hierarchical approach, where the SoC is typically partitioned into smaller sub-systems, and sub-systems are further partitioned into smaller functional modules, that can be designed efficiently by the EDA tools.  At each of these hierarchies (SoC and sub-system), the floorplanning step aims to determine the optimal physical locations and shapes of the individual blocks constituting the SoC or a subsystem. 
The outputs from the partitioning step are partition-level area budgets, partition-level connectivity information, connectivity to external terminals, and the locations of external terminals. These serve as the input to the floorplanning problem. In addition, the floorplanning problem typically has several constraints that govern the legal positions of some of the blocks. Floorplanning stands as a pivotal step in the IC design cycle, serving as a crucial bridge connecting the front-end and back-end phases of the design process. It gives early feedback on the architectural choices and the output of the floorplanning phase, such as the placement of blocks and their shapes, impacts downstream physical design targets. \\

\noindent At a high-level, the floorplanning formulation is a multi-objective optimization problem involving assignment of locations and shapes to individual blocks, while considering various design constraints. It is typically characterized by the following inputs, constraints and outputs.\\

 \textbf{Inputs:} Given:
 \begin{itemize}
     \item a set  $B = \{b_1, b_2, ..., b_k\}$, comprising  $k$  circuit blocks (we also interchangeably refer to them as modules or blocks).
     \item a set $A = \{a_1, a_2, ..., a_k\}$, representing the estimated post-routed layout area budgets of each partition.
     \item a set $T = \{ t_1, t_2, ..., t_r\}$, representing $r$ terminals for external interfacing of the floorplan system. Each terminal is a point in 2-D space (typically on the periphery of the system).
     \item a circuit hyper-graph $G$ that captures the connectivity information which is comprised of hyper-edges $ E = \{e_1, e_2,... e_n \}$, representing $n$ hyper-nets to capture block-to-block and block-to-terminal connections. Hyper-edge $e_i$ is the set of blocks and/or terminals that are connected together through net $i$. 
 \end{itemize}

 \textbf{Constraints:} 
 \begin{itemize}
     \item Outline constraints: Modern ASIC design relies on a hierarchical (top-down) floorplanning style, with a "fixed-outline" or "fixed-die" requirement for the SoC and its sub-systems. A floorplan optimized for area without accounting for the fixed-outline constraint may prove impractical, as it may fail to fit within the specified outline.
     \item Boundary constraints: These indicate that a partition must align with a specific edge or corner of the floorplanning outline. This alignment is necessitated by external interfacing requirements of the blocks or system-level thermal considerations. 
     \item Grouping constraints: These require that some blocks must be physically abutted. This is particularly pertinent for blocks operating on the same voltage or those that need to be simultaneously  powered off.
     \item Pre-placement constraints: This captures previously-established optimal placement locations of blocks that are reused from legacy designs. Furthermore, throughout the physical design phase, as partition-level area budgets become more refined due to stable RTLs and feedback from the physical design flow, certain blocks may undergo reshaping, while stable blocks may remain unaffected. Such stable blocks become part of the pre-placement constraints.
 \end{itemize}

\textbf{Objective:} Assignment of overlap-free block locations, $L = \{ (x_i, y_i) | i = 1, 2, ... k\}$, and block aspect ratios, $R = \{ar_{1}, ar_{2}, ..., ar_{k} \}$, such that all constraints are satisfied and the following multi-objective cost metric is minimized:
\begin{equation}
  \label{eq:classical_sa}
   Cost = \alpha * HPWL + \beta * area.
\end{equation}
Here, $area$ is the area of the bounding box of all placed blocks and $HPWL$ is the half-perimeter wire-length. $\alpha$ and $\beta$ are the HPWL and area cost weights, respectively. HPWL is defined as follows: let $\{(x_{i,1},y_{i,1}),\ldots,(x_{i,n_i},y_{i,n_i})\}$ be the center points of the blocks and/or the locations of the terminals that belong to net $i$, HPWL is given by:
\begin{equation}
  HPWL = \frac{1}{2} \sum\limits_{i=1}^{n} \max\limits_{u,v<n_i} |x_{i,v} - x_{i,u}| + \max\limits_{u,v<n_i} |y_{i,v} - y_{i,u}|,
\end{equation}
i.e, the cost for each net is half the perimeter of the rectangle encompassing the net's endpoints.


\section{Related Work}
Modern physical design flows solve the floorplanning problem using a top-down fixed-outline \cite{kahng20} methodology, in which the outline of the floorplan system is predetermined. A fixed-outline constraint \cite{adya03} is more challenging than variable-die or outline-free floorplanning, due to a more restrictive solution space. A good floorplan representation \cite{fprep05, nonslice23} is crucial to reduce the search space and enable faster convergence to an optimal solution.The goal of a good representation is to encode the relative positions and shapes in a way that is amenable to perturbation. Some popular topological overlap-free floorplan representations are: Normalized Polish Expression \cite{npe86}, Corner Block List \cite{cbl00}, O-tree \cite{otree00}, \texttt{B*-tree} \cite{btree06}, Corner Sequence \cite{cbl00}, Sequence Pair \cite{spair98}, Transitive Closure Graph \cite{tcg01}, Transitive Closure Graph - Sequence Pair \cite{tcgs02}, and Adjacent Constraint Graph \cite{acg04}. A comprehensive summary of the trade-off between various representations are summarized in \cite{fprep05}. Computationally, floorplanning is an NP-hard problem. As the number of blocks grows, the search space tends to grow exponentially. Prior works for floorplan optimization can be categorized into analytical, heuristics-based, and learning-based approaches. \\

\begin{table*}[h]
\centering
\caption{Summary of previous floorplanning formulations along with the constraints they support.}
\begin{tabular}{c c c c c c c}
\toprule

Reference  & Fixed-outline & Pre-placed & Boundary & Grouping &    Comments   \\\hline
\hline

\cite{adya01, adya03}  & \cmark & \xmark & \xmark & \xmark  & SA-based \\
\cite{young98}  & \cmark & \cmark & \xmark & \xmark   & SA-based\\
\cite{young99, jianbang01, yuchun01}  & \cmark & \xmark & \cmark & \xmark   & SA-based\\
\cite{learnfp20, gfp22, kdd22, defer08}  & \cmark & \xmark & \xmark & \xmark   & Learning-based\\
\cite{ga00, ga12}  & \cmark & \xmark & \xmark & \xmark   & GA-based \\
\cite{ant09,  ant13}  & \cmark & \xmark & \xmark & \xmark   & ACO-based \\
\cite{antsa16, hybrid23}  & \cmark & \xmark & \xmark & \xmark   & Hybrid-based \\
\cite{lp03, milp91}  & \cmark & \xmark & \xmark & \xmark   & LP-based analytical\\
\cite{quasi21, poisson23, convex08, sun24, sachin06}  & \cmark & \xmark & \xmark & \xmark   & Two-stage analytical\\
\cite{btree06, effsa12}  & \cmark & \cmark & \cmark & \xmark   & Sub-optimal packing and simpler testcases\\
\cite{young04}$^{*}$  & \cmark & \cmark & \cmark & \cmark   & Soft-penalty and sub-optimal results\\
Our work  & \cmark & \cmark & \cmark & \cmark   &  Hard constraints and Better results\\
\hline
\end{tabular}
\label{tab:prev_works} 
\end{table*}

\noindent \textbf{Analytical \& Heuristic Solvers:} While analytical solvers that use linear programming, \cite{lp03} and mixed-integer linear programming \cite{milp91} aim to solve the exact floorplanning problem using a closed-form solution, such methods are not scalable as the number of blocks grows. As an alternative, two-stage analytical solutions \cite{quasi21, poisson23, convex08, sun24, sachin06} propose a continuous-space coarse floorplanning in the first stage, followed by local overlap removal and shape refinement in the second stage. Since the second stage is usually limited to local refinement, the global constraint-awareness needs to be incorporated into the coarse stage. Therefore, they suffer from the inability to guarantee violations-free start points for the second stage, particularly when dealing with complex pre-placed or boundary constraints. In the class of heuristic optimization, the most common techniques are simulated annealing \cite{adya03, btree06, effsa12}, genetic algorithms \cite{ga12, ga00}, ant-colony optimization \cite{ant09, ant13}, and hybrid methods that combine multiple heuristic approaches \cite{antsa16, hybrid23}. \\

\noindent \textbf{Learning-based solvers:} One emerging category of floorplanning techniques are learning-based algorithms. Efficient floorplan representations \cite{fprep05} establish a foundation for formulating the floorplanning problem as a learning-to-search mechanism. He et al. \cite{learnfp20} use sequence-pair representations and iteratively search over these representations using a deep-Q algorithm. \textit{GoodFloorplan} \cite{gfp22} combines graph convolution and a sequence-pair formulation to capture the connectivity and topological information of the layout, and employs an actor-critic network to iteratively search for better neighboring floorplans. Amini et al. \cite{kdd22} use a deep reinforcement learning (RL) agent to perform floorplanning, using the corner-block representation. They focus on wire-length optimization using RL-PPO for one-shot optimal corner-block construction. However, the absence of  incremental floorplan refinement steps is a major weakness of such one-shot floorplanners.
\textit{Defer} \cite{defer08} provides a non-stochastic method for compacting a slicing floorplan, and generates a non-slicing floorplan. By adopting the principle of Deferred Decision Making (DDM) during the node-combining step in generalized slicing trees, the slicing tree in \textit{Defer} represents a large number of slicing floorplan solutions.  Since deferring ensures that all of these solutions are stored by a single shape curve, they choose a good slicing floorplan that fits into the fixed outline. All of the above methods, however, do not consider complex instances of the floorplanning problem that include pre-placed blocks, boundary, and clustering constraints. The lack of reproducible code and model architecture details makes it difficult to validate the claim that these learning-based methods outperform traditional methods like simulated annealing. \\

\noindent \textbf{Modern Floorplanning Constraints:} Modern industry-relevant floorplanning tasks generally include some hard placement constraints \cite{young04} such as boundary (edge and corner), pre-placed blocks, and grouping constraints. 
One of the primary drawbacks of existing floorplanning methods is the absence of mechanisms for automatically incorporating hard constraints. While these constraints can be represented by extra terms in the cost function, we show that this 'soft' representation of the constraints does not work in practice. 
\cite{adya01} proposed methods to improve the search efficiency, but that was restricted to fixed-outline constraints alone. Chang et al. propose a pre-placement-aware \texttt{Btree} formulation \cite{btree06}. The proposed method, however, leads to large white-space, thereby violating the fixed-outline constraint. Young et al. handle pre-placement constraints \cite{young98} by focusing on slicing representations but their approach also leads to sub-optimal white-space.  Murata et al. incorporate pre-placement constraints using a sequence-pair representation, but also suffer from inefficient packing around the pre-placed blocks. Although a number of previous works \cite{jianbang01, yuchun01, young99} include boundary constraints in their formulation, they either fix the violations in a post-processing step or use a penalty term that makes it a soft constraint. We could find only one work \cite{young04} that comprehensively covers all placement constraints using a simulated annealing approach on sequence-pair representations. They augment the constraint graphs used in the traditional sequence-pair representation with constraining edges that indicate placement constraints.  This is done by incorporating an additional penalty cost for infeasible solutions. Such an approach needs careful balancing of the penalty cost and leads to sub-optimal white-space, as indicated in their results. 
        
\section{Methods}
\subsection{Preliminaries}
Before introducing our proposed \textit{PARSAC} framework in Section ~\ref{sec:parsac}, we first review the \texttt{B*-tree} representation, its benefits and its solution space.
\texttt{B*-trees} \cite{btree06} are ordered binary trees, augmented with rules that allow them to model compact floorplans. By definition \cite{btree06}, in a compact floorplan, no block can be moved towards the bottom or to the left of the floorplan without overlapping another block. This class of area-optimal floorplans always has an equivalent \texttt{B*-tree} representation. \texttt{B*-trees} are easy to implement and manipulate using primitive tree operations such as search, insertion, and deletion which can be done in constant, constant, and linear time, respectively.

\begin{figure}
    \centering
\includegraphics[scale=0.25]{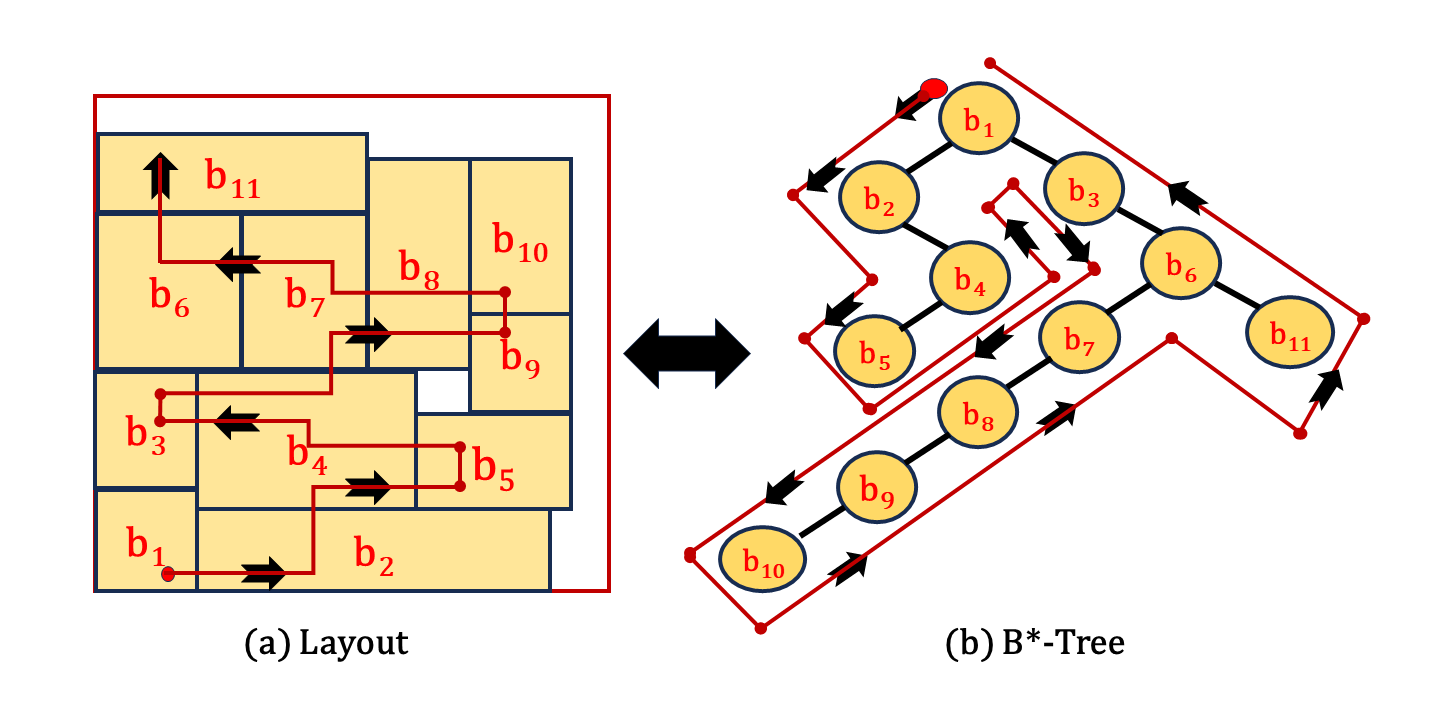}
    \caption{An example of a compacted floorplan (left) and the corresponding \texttt{B*-tree} (right) representation. Given a compact layout, a tree representation can be obtained by traversing the blocks, starting from the lower-left module $b1$. Likewise, given a \texttt{B*Tree}, layout can be obtained by a DFS traversal, starting from the root node $b1$.}
    \label{fig:btree}
\end{figure}

\noindent \textbf{Layout-to-Tree Translation:} As shown in Fig.\ref{fig:btree}, the translation operation between the floorplan layout and the \texttt{B*-tree} representation can be performed in linear time by using a recursive procedure similar to the depth first search (DFS) algorithm. In the \texttt{B*-tree}, each node denotes a module, with the root representing the module on the bottom-left corner of the layout i.e., $(x_1, y_1) = (0, 0)$. The edges of the tree denote a parent-child relation that indicates the relative placement of two nodes. For example, the left child $b_{j1}$ of a parent node $b_i$ denotes the module that is placed at the lowest possible position immediately to the right of the parent node i.e., $x_{j1} = x_i + w_i$, where $w_i$ is the width of module $b_i$. The right child $b_{j2}$ of a parent node $b_i$ denotes the module that is placed at the lowest possible position above $b_i$ i.e., $x_{j2} = x_i$. The $y$-coordinates of the modules can be  computed in constant time, using a doubly-linked horizontal contour data structure\cite{btree06}. Similarly, given a layout, the associated \texttt{B*-tree} can be obtained by traversing the layout starting from the lower-left module. 

\noindent \textbf{Search Space:} For a floorplan problem with $k$ modules, the size of the solution space of possible \texttt{B*-trees} in an unconstrained settings is given by:
$$O(k ! * \frac{2^{2k}}{k^{1.5}} )$$
For an SoC with $120$ blocks (typical size of modern SoCs), there are approximately $10^{250}$ possible solutions. Search methods traverse this space by iteratively modifying the tree using the following three primitive operations:
\begin{itemize}
    \item Swap two nodes.
    \item Delete a node and insert it elsewhere.
    \item Modify the shape (or aspect ratio) of a node.
\end{itemize}


\subsection{Constraints-aware simulated annealing}
\label{sec:parsac}
\begin{figure}
    \centering
\includegraphics[scale=0.45]{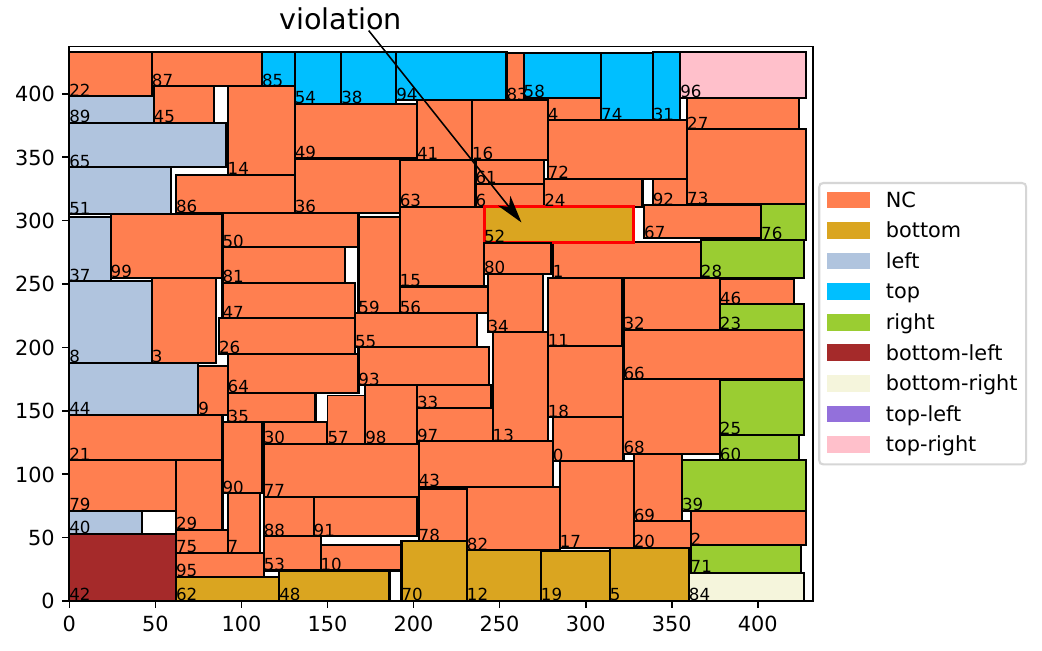}
    \caption{An example of a hard constraints local minimum. There is one violating block (shown with a red outline) that should be placed on the bottom boundary. The bottom boundary is already filled with constrained blocks. Moving the violating block and placing it to the right of one of the bottom-boundary blocks will increase the width of the bottom row and move the right edge of the floorplan. This will result in constraint violations at all the constrained right edge blocks (green blocks) because they will no longer be at the right edge of the floorplan.}
    \label{fig:stuck_sa}
\end{figure}

Traditional SA seeks to minimize area and wire-length (the cost function defined in Eq.~\ref{eq:classical_sa}). It does not consider hard placement constraints such as boundary constraints (some blocks have to be placed at a particular edge or corner of the floorplan), or pre-placement constraints (some blocks have to be placed at a specific location). One straightforward way to make SA aware of these constraints is to include them in the cost function by modifying Eq.~\ref{eq:classical_sa} to be:
\begin{equation}
  \label{eq:naive_sa}
   Cost = \alpha * HPWL + \beta * area  + \gamma * number\_of\_violated\_constraints,
\end{equation}
where $\gamma$ is the weight attached to violated constraints. We show, however, that SA typically fails to satisfy boundary constraints and pre-placement constraints, even for large values of $\gamma$. The failures of SA to meet these constraints are typically because SA gets stuck in configurations with violated placement constraints, but where there is no single move that would decrease the number of violated constraints. This situation is illustrated in Fig.~\ref{fig:stuck_sa}. We call such a configuration a `hard constraints local minimum'. Such local minima are even harder to escape for large values of $\gamma$.

The `hard constraints local minimum' problem cannot be solved by tuning $\gamma$. A larger $\gamma$ would make the probability of escaping the local minimum harder, while a smaller $\gamma$ would lead to final solutions where SA decided to violate some hard constraints in favor of reducing  $HPWL$ or $area$. Our core observation is that we cannot depend on traditional SA mechanisms alone to converge to configurations that satisfy the hard placement constraints.

We augment SA with two novel techniques to allow it to perform well in the presence of boundary and pre-placement constraints: constraints-fixing moves and \texttt{B*-tree} with anchored blocks.

\subsubsection{Constraints-fixing moves}
To solve the `hard constraints local minimum' issue, we augment the standard SA move set (swapping or moving blocks in the \texttt{B*-tree} or aspect ratio adjustment) with moves whose sole purpose is to fix violated boundary constraints. We call these moves constraints-fixing moves. A constraints fixing move is always accepted, irrespective of how it changes the cost function. At every SA step, we either try a standard SA move and accept it or reject it based on the resulting cost delta, or we make a constraints-fixing move. 

Constraints-fixing moves have a low probability of being applied at an SA step ($\approx 0.0005$). Algorithm~1 describes our modified SA move set. Most of the time, we apply a standard SA move such as swapping, moving, or perturbing the aspect ratio. Every once in a while, and if there are violated boundary constraints, we explicitly pick a block that violates the boundary constraints and fix its constraint violation. We do that in one of two ways:
\begin{enumerate}
\item If there are blocks at the required boundary (for example left boundary) that are unconstrained - i.e, these blocks do not have to be at that boundary - then we swap the violating block with one of these unconstrained blocks.
\item If all the blocks at the required boundary are constrained - i.e, they have to be at that boundary - then we move the violating block and make it the right child (to be placed on top) or the left child (to be placed to the right) of one of the constrained blocks at the required boundary. 
\end{enumerate}

\begin{algorithm}
  \caption{The CA-SA move set}
\begin{algorithmic}
\label{alg:CA_SA}
\State {\bf Inputs}: $B$: B*Tree
\State \# $rand()$ produces random numbers uniformly in [0,1]
  \If {rand() > 0.0005 or boundary\_constraints\_satisfied()} 
  \If{rand() > 0.333}  \Comment {Aspect-ratio adjustment}
  \State $b_1 \leftarrow pick\_random\_block()$
  \State $Perturb\_AR(b_1)$
  \Else \Comment {Move or swap blocks}
  \State $b_1,b_2 \leftarrow pick\_random\_block\_pair()$
  \If {rand() > 0.5} \Comment {Swap two blocks}
  \State $B.swap(b_1,b_2)$
  \Else \Comment{Move $b_1$ to be child of $b_2$}
  \State $B.remove(b_1)$
  \If {rand() > 0.5} 
  \State $B.add(b_1,b_2,LEFT)$\Comment{make $b_1$ left child of $b_2$}
  \Else
  \State $B.add(b_1,b_2,RIGHT)$\Comment{make $b_1$ right child of $b_2$}
  \EndIf
  
  \EndIf
  
  \EndIf
  \Else \Comment{move for fixing boundary constraints violations}
  \State $b \leftarrow pick\_random\_violating\_block()$
  \State $S_u \leftarrow unconstrained\_blocks\_at\_loc(b.required\_loc)$
  \If{$S_u \neq \Phi$} \Comment {$S_u$ is not empty}
  \State $b_s \leftarrow random\_sample(S_u)$
  \State $B.swap(b,b_s)$ \Comment{Swap with block at the required boundary}
  \Else \Comment{all blocks at required boundary are constrained}
  \State $S_c \leftarrow constrained\_blocks\_at\_loc(b.required\_loc)$
  \State $b_s \leftarrow random\_sample(S_c)$
  \State $B.remove(b)$
  \If{$b.required\_loc \in \{left,right\}$}
  \State $B.add(b,b_s,RIGHT)$ \Comment{make $b$ right child of $b_s$}
  \Else \Comment{$b.required\_loc$ is top or bottom}
  \State $B.add(b,b_s,LEFT)$  \Comment{make $b$ left child of $b_s$}
  \EndIf      
  \EndIf
  \EndIf  

\end{algorithmic}
\end{algorithm}

\subsubsection{\texttt{B*-tree} with anchored blocks}
In the presence of pre-placed blocks that have to be placed at specific $x$-$y$ locations, we can modify the SA cost function to penalize the placement of these blocks away from the required $x$-$y$ locations. We show, however, that this usually does not guarantee exact placements at the required location. To tackle these pre-placement constraints, we modify the way the \texttt{B*-tree} is represented and the way the floorplan is constructed from the \texttt{B*-tree}. 

We modify the \texttt{B*-tree} so that the blocks with pre-placement constraints ignore the position dictated by the \texttt{B*-tree}. Instead, they are anchored to the pre-placed positions. The children of these anchored blocks are still placed relative to these anchored blocks according to the positions dictated by the \texttt{B*-tree}. As shown in Fig.~\ref{fig:btree}, the \texttt{B*-tree} is traversed in a left-first, depth-first manner to sequentially place all the blocks. Algorithm~2 outlines the steps for placing one block from the B*Tree. The pre-placed constraints are thus satisfied by construction.

\begin{algorithm}
  \caption{Block placement step}
\begin{algorithmic}
\label{alg:anchored_blocks}
\State {\bf Inputs}: $b$: block to be placed
  \If {$is\_preplaced(b)$} \Comment{Block has preplacement constraint}
  \State $b.x,b.y \leftarrow b.preplaced\_x,b.preplaced\_y$
  \Else
  \If {$b.parent == None$} \Comment{$b$ is root}
  \State $b.x,b.y \leftarrow 0,0$
  \ElsIf {$b.parent.left\_child == b$}
  \State $b.x \leftarrow b.parent.x + b.parent.width$
  \Else \Comment{$b$ is a right child}
  \State $b.x \leftarrow b.parent.x$
  \EndIf
  \State $b.y \leftarrow compact(b)$ \Comment{Place $b$ at the lowest possible position}
  \EndIf

\end{algorithmic}
\end{algorithm}

\subsubsection{Fixed outline and grouping constraints}
We incorporate these two constraint categories into our CA-SA pipeline in a relatively straightforward manner. For the outline constraint, we simply add an extra cost term in the CA-SA cost functions:
\begin{equation}
  cost_{outline} = \eta\left([bbox.W - outline.W]^+ + [bbox.H - outline.H]^+\right),
\end{equation}
where $bbox.W$ and $bbox.H$ are the width and height of the bounding box of the placed blocks. $outline.W$ and $outline.H$ are the width and height of the fixed outline in which the blocks need to fit. $[x]^+ \equiv max(0,x)$ is the rectification function and $\eta$ is the fixed outline cost coefficient.

We also represent the grouping constraint as an extra term in the cost function. Let $\{g_1,\ldots,g_G\}$ be the required block groupings, where $g_i$ is the set of blocks that belong to group $i$. In an actual floorplan, the blocks in $g_i$ will form $z_i$ clusters. The grouping constraint for  $g_i$ is satisfied if $z_i=1$ The grouping cost is given by:
\begin{equation}
  cost_{grouping} = \zeta\sum\limits_{i=1}^G \frac{(z_i - 1)}{|g_i|},
\end{equation}
where $\zeta$ is the grouping cost coefficient.

\subsubsection{Multi-instantiation and shape constraints}
Our CA-SA also supports multi-instantiation constraints. A multi-instantiation constraint requires some blocks to always have the same aspect ratio since these blocks are instances of the same master. We support these constraints by always identically perturbing the aspect ratios of these blocks during the search. Shape constraints limit the minimum and maximum aspect ratios of blocks. We support them by always clamping aspect ratios back to these limits if they are pushed beyond them in an SA step. We do not provide results for these two constraint types as they are straightforward to support in a manner that guarantees correctness.

\subsection{PARSAC (Parallel Simulated Annealing with Constraints)}
\begin{figure}
    \centering
\includegraphics[scale=0.4]{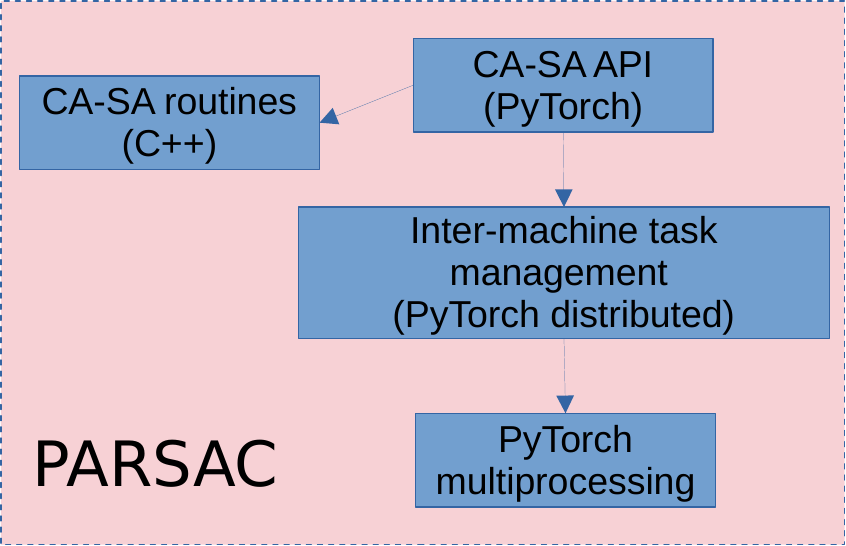}
    \caption{PARSAC architecture.}
    \label{fig:parsac_arch}
\end{figure}

In order to improve the quality of the solutions produced by our constraints-aware SA pipeline, we developed a parallelization framework that allows us to effectively scale to an arbitrary number of machines and CPU cores. This framework, PARSAC, is built on top of `PyTorch' and `PyTorch Distributed'. The general architecture of PARSAC is illustrated in Fig.~\ref{fig:parsac_arch}. The core routines of CA-SA are implemented in C++. We create a PyTorch-based Python API that interfaces to the C++ code. There are two levels of parallelization: inter-machine parallelization which is implemented using the PyTorch distributed framework and intra-machine parallelization using the multiprocessing module.

We use a naive parallel search where we instantiate one CA-SA worker per each CPU core. For $M$ machines with $C$ cores each, we will have $M*C$ workers. Each CA-SA worker runs independently. We collect the final $M*C$ solutions and use these to construct Pareto optimal fronts with trade-offs between area and wire-length. 

\section{Results}
\subsection{Legacy floorplan problems}
We first test PARSAC on classical floorplanning problems from the literature with no pre-placement constraints. We compare against prior work based on simulated annealing as well as against more recent RL-based techniques. These previous methods assumed hard blocks, i.e, they did not allow adjustments to the blocks' aspect ratios. We present two sets of PARSAC results: PARSAC(soft) which performs aspect ratio adjustments and PARSAC(hard) which does not. In the PARSAC(soft) results we only allow a block's aspect ratio to vary in the range $[\frac{1}{3},3]$.

We test on the \texttt{n100}, \texttt{n200}, and \texttt{n300} floorplans which have $100$, $200$, and $300$ blocks, respectively. These are fixed outline floorplanning problems where the goal is to minimize wire-length while placing all blocks within the outline. The outline is square and is $10\%$ larger than the total block area. We do not scale the terminal locations but keep all the terminals at their problem-defined positions, which is the convention adopted by prior work. We run $112$ PARSAC workers in parallel across $112$ CPU cores. Each worker runs for $10$ million SA steps. We report results from the two workers with the best and worst wire-lengths. All workers are able to place the blocks to fit within the fixed outline.  The results are shown in table~\ref{tab:legacy_results}. PARSAC(hard) outperforms previous baselines and achieves significantly lower wire-length. PARSAC(soft) achieves even lower wire-length, which shows that PARSAC can effectively exploit the extra degrees of freedom afforded by soft blocks. We also report PARSAC's run-time. The experiments were run on a $2$-socket fifth generation Xeon Scalable server processor (Emerald Rapids) with 56 cores per socket.

\begin{table*}
\centering
\caption{Wire-length (mm) for fixed outline floorplanning. For PARSAC, we report the worst and best results across $112$ workers.}
\begin{tabular}{|l||c|| c|c|c|c|c|c|c|c||c|}\hline
Design     & \begin{tabular}{@{}c@{}}PARSAC \\ run-time(mins)\end{tabular} & PARSAC(soft) & PARSAC(hard)   & Parquet~\cite{parquet} & B*-FOPP~\cite{btree06} & SP-FOPP~\cite{antsa16} & GoodFloorPlan~\cite{gfp22} \\ \hline \hline
n100     & 9.3 & {\bf 288}-301 & {\bf 291}-310  & 344 & 323 & 319 & 309  \\ \hline
n200     & 31.2 & {\bf 526}-553 & {\bf 530}-557  &654 & 599 & 575 & 558  \\ \hline
n300     & 67.8 & {\bf 627}-663 & {\bf 628}-673  & 799 & 757 & 705 & 691  \\ \hline
\end{tabular}
\label{tab:legacy_results}
\end{table*}

\subsection{Floorplan problems with placement constraints}
\begin{figure*}[!bh]
  \centering
  \begin{subfigure}{0.3\textwidth}
    \includegraphics[width = \textwidth]{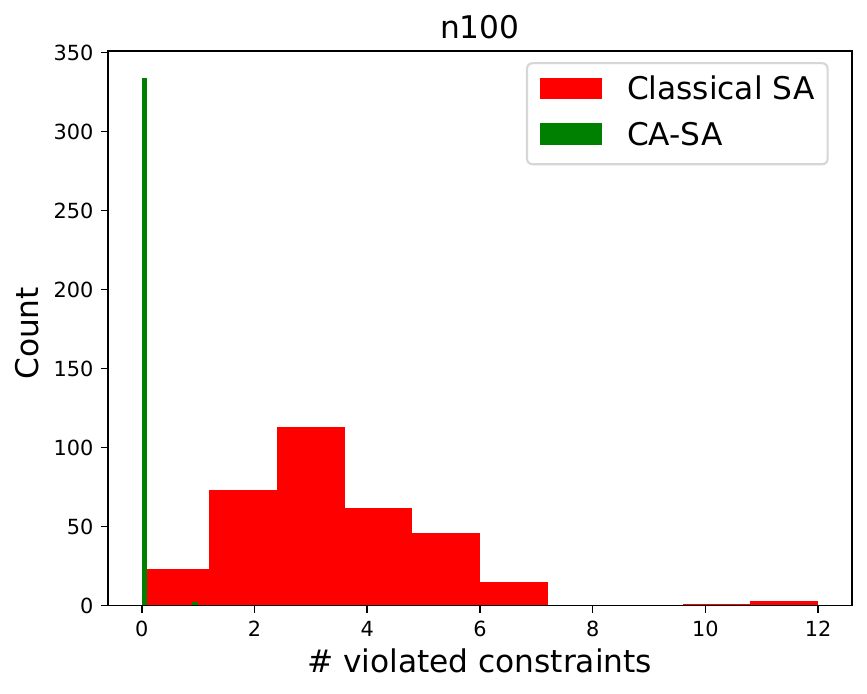} 
    \subcaption{}
  \end{subfigure}
  \quad
  \begin{subfigure}{0.3\textwidth}
    \includegraphics[width = \textwidth]{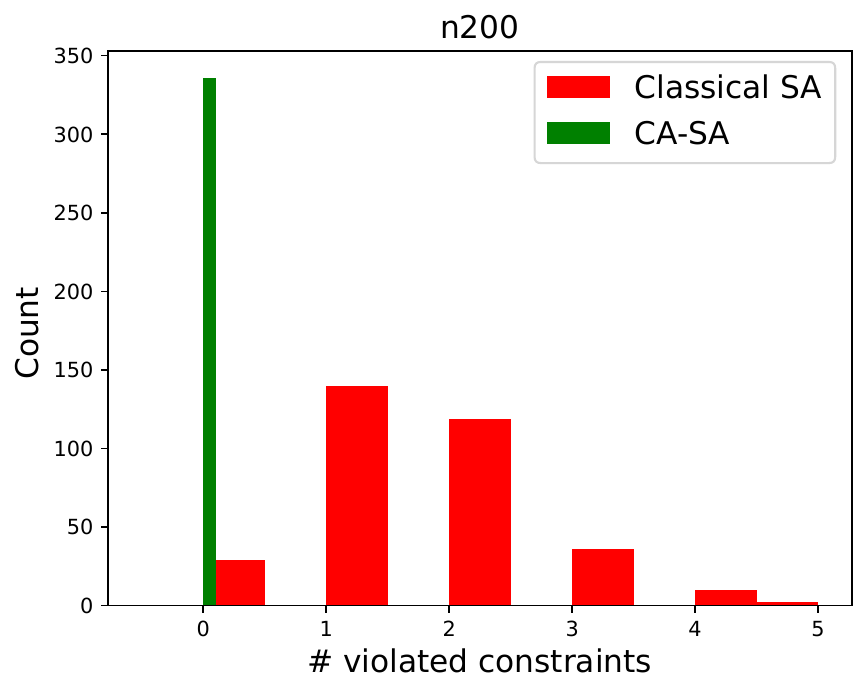} 
    \subcaption{}
  \end{subfigure}
  \quad
  \begin{subfigure}{0.3\textwidth}
    \includegraphics[width = \textwidth]{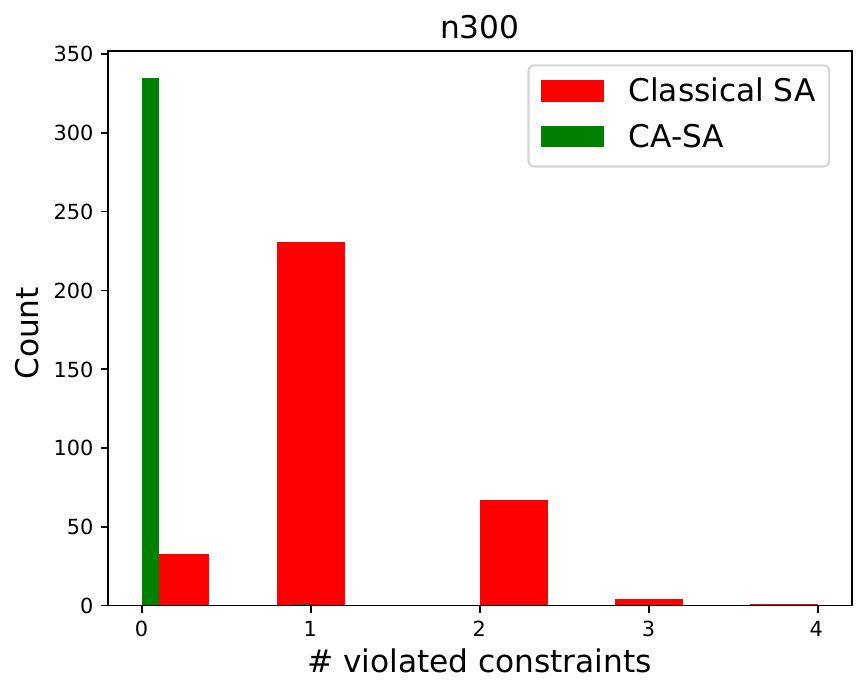} 
    \subcaption{}
  \end{subfigure}

  \caption{Histogram of boundary constraints violations for CA-SA and classical SA for the constrained n100, n200, and n300 problems. The statistics for each problem were collected from 336 trials.}
  \label{fig:violations_histogram}
\end{figure*}

The popular floorplanning benchmarks in the literature do not have hard preplacement constraints. To evaluate the effectiveness of PARSAC, we extend the \texttt{n100}, \texttt{n200}, and \texttt{n300} problems with hard pre-placement constraints as follows:
\begin{enumerate}
\item Randomly pick 4 groups of 7 blocks each. Assign to each group a left boundary, right boundary, top boundary, or bottom boundary constraint.
\item Randomly pick 4 blocks. Assign to each block a bottom-right, bottom-left, top-right, or top-left boundary constraint.
\item Randomly pick 3 groups of 3 blocks each. Assign a grouping constraint to all blocks within the same group.
\item Randomly pick 10 blocks and assign to them a hard pre-placement constraint at random $x$-$y$ locations within the fixed outline.
\end{enumerate}
The random block picks are non-overlapping. We compare CA-SA against classical SA. For classical SA, we support the different constraints through extra terms in the cost function as illustrated in  Eq.~\ref{eq:naive_sa}. For each augmented benchmark, we run $336$ trials for CA-SA and for classical SA. We vary the cost coefficients across trials. Fig~\ref{fig:violations_histogram} plots the frequency of boundary constraints violations for CA-SA and classical SA. We see that without the constraints-fixing move in SA, the resulting floorplan almost always has some boundary constraints violation. CA-SA, with its constraints-fixing moves, virtually guarantees no boundary constraints violations.

\begin{figure*}[h]
  \centering
  \begin{subfigure}{0.45\textwidth}
    \includegraphics[width = \textwidth]{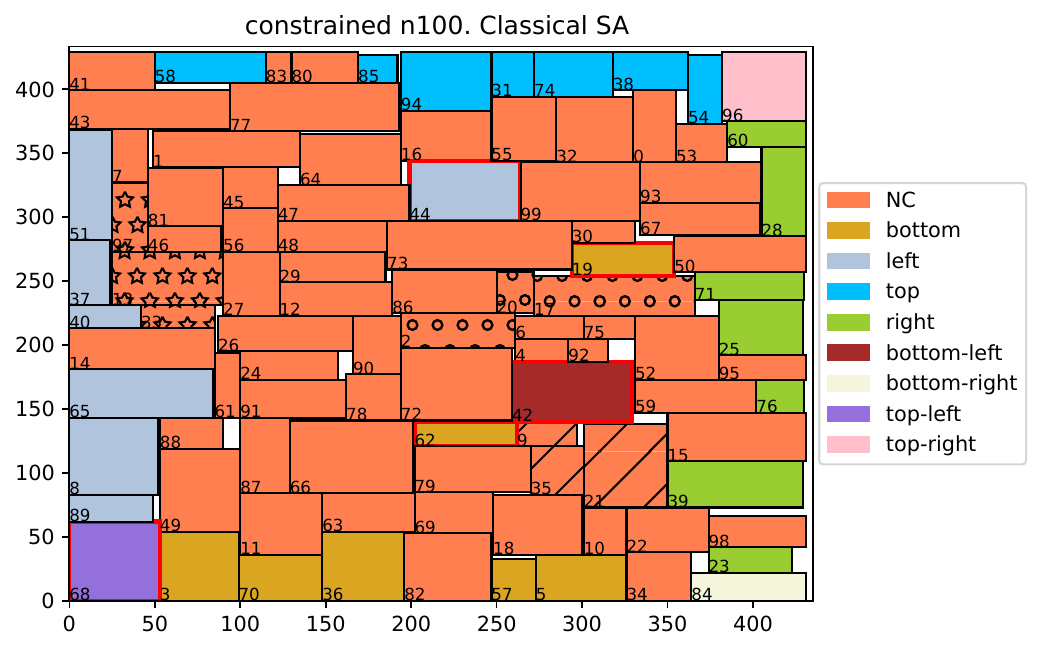} 
    \subcaption{}
    \label{fig:vis_a}
  \end{subfigure}
  \begin{subfigure}{0.45\textwidth}
    \includegraphics[width = \textwidth]{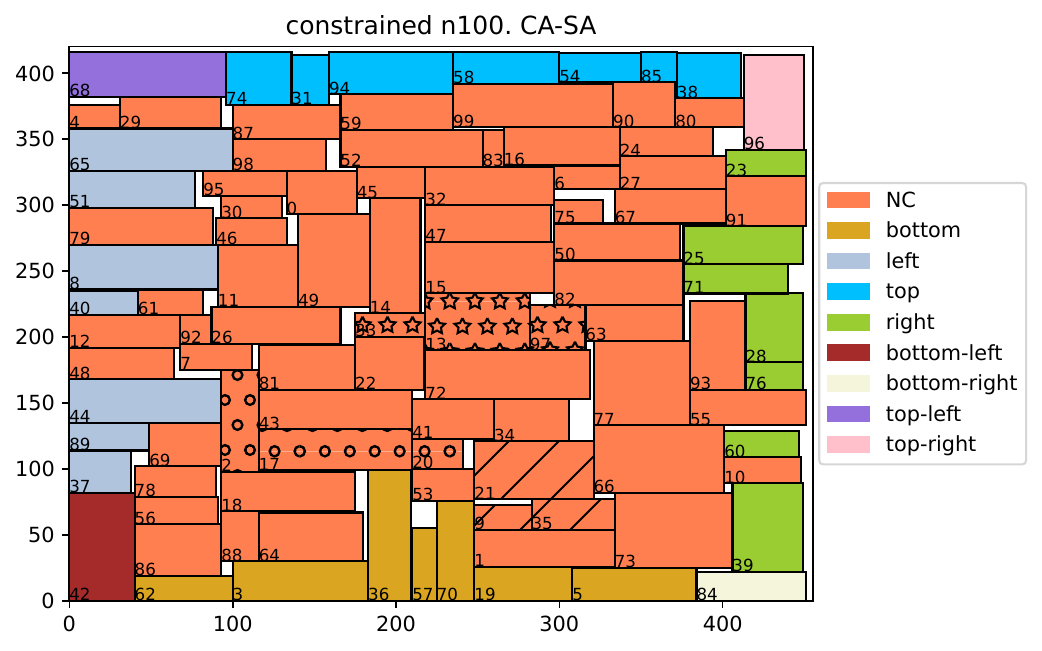} 
    \subcaption{}
    \label{fig:vis_b}    
  \end{subfigure}
  \caption{(\subref{fig:vis_a}) Sample floorplan from classical SA for the constrained \texttt{n100} problem. There are five blocks that violate their boundary constraints. The violating blocks have a red outline. Blocks that are part of the same grouping constraint have the same fill pattern (circle, star, or cross-hatch patterns). (\subref{fig:vis_b}) Sample floorplan from CA-SA for the same problem as \subref{fig:vis_a}. All blocks satisfy the boundary constraints.}
  \label{fig:violation_visualization}
\end{figure*}

Figure~\ref{fig:violation_visualization} shows two floorplans for the randomly constrained \texttt{n100} problems. Both CA-SA and classical SA are generally able to satisfy the grouping constraints solely by including the number of violated grouping constraints in the cost function. The floorplan found by classical SA, however, has five boundary constraints violations while the floorplan from CA-SA has no boundary constraints violations. 

\begin{figure*}
    \centering
\includegraphics[scale=0.5]{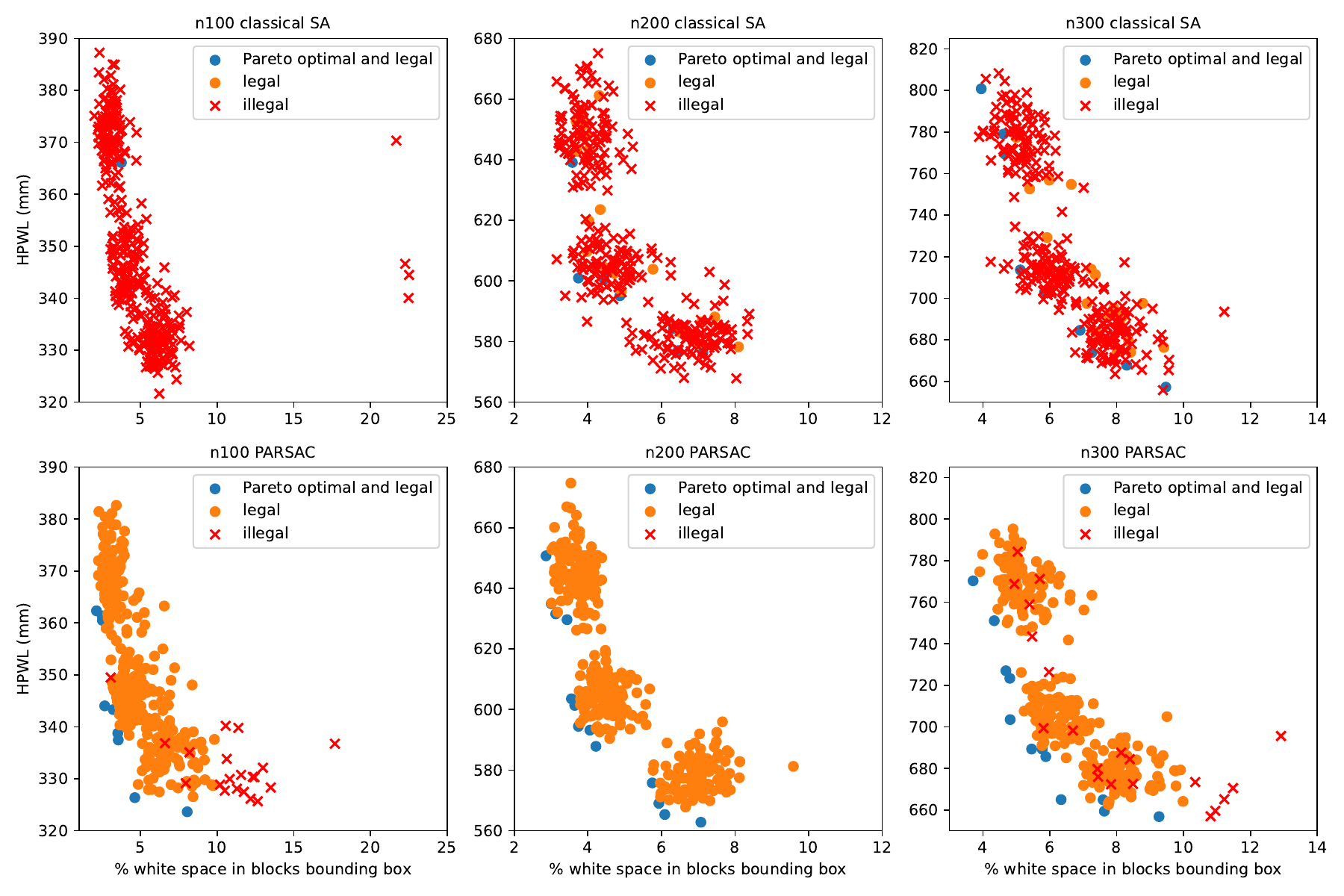}
    \caption{The distribution of $336$ solution points for the constrained \texttt{n100}, \texttt{n200}, and \texttt{n300} problems for classical SA and PARSAC. We quantify the solution quality using the HPWL and the percentage of white-space in the block's bounding box (the minimal box that contains all the blocks). The top rows shows the results for classical SA. Most of the floorplans are illegal. The bottom row shows the results for PARSAC which finds legal floorplans most of the time.}
    \label{fig:pareto}
\end{figure*}

The inability of classical SA to reliably satisfy the hard boundary constraints makes it very difficult to generate a rich {\bf legal} Pareto front. That is because most of the discovered floorplans would be illegal. Figure~\ref{fig:pareto} illustrates this issue: Across 336 trials, classical SA mostly generates illegal solutions for the constrained \texttt{n100}, \texttt{n200}, and \texttt{n300} problems. A floorplan can be illegal due to boundary constraints violations, grouping constraints violations, or fixed outline violations. The pre-dominant type of violations for classical SA is boundary constraints violations. For the same number of trials, however, PARSAC can generate mostly legal floorplans leading to a richer legal Pareto front. The pre-dominant violation types in PARSAC are grouping and fixed outline violations. PARSAC is an ideal choice for exploring the Pareto-optimal trade-offs between design metrics (such as area and wire-length) for constrained floorplanning problems due to its massively parallel nature and because it mostly generates legal floorplans.

The straightforward approach to handle hard pre-placement constraints in classical SA is to modify the SA's cost function to penalize placing pre-placed blocks away from their target locations. Let $B = {b_1,\ldots,b_P}$ be the set of blocks with hard pre-placement constraints. We require block $b_i$ to be placed at position $(x_i,y_i)$. The extra term in the classical SA's cost function has the form:
\begin{equation}
  cost_{preplaced} = \theta\sum\limits_{i=1}^P |(b_i.x - x_i)| + |(b_i.y - y_i)|,
\end{equation}
where $\theta$ is the pre-placed cost coefficient and $(b_i.x,b_i.y)$ is the location of block $b_i$ found by SA. We observe that this additional cost term does indeed move blocks with pre-placement constraints close to the required location. However, the placement is practically never exact. Table~\ref{tab:placement_delta} shows the mean deviations along the $x$ and $y$ dimensions for the pre-placed blocks away from their pre-placed locations. For PARSAC, the deviations are zero by construction since PARSAC uses a \texttt{B*-tree} with anchored blocks to guarantee that pre-placed constraints are satisfied. 

\begin{table}
\centering
\captionsetup{justification=centering}
\caption{Mean placement deviations for pre-placed blocks with classical SA.}
\begin{tabular}{|l||c|c|}\hline
Design     & x & y  \\ \hline \hline
n100     & 2.79 & 0.80   \\ \hline
n200     & 3.29 & 0.37   \\ \hline
n300     & 3.31 & 0.44   \\ \hline
\end{tabular}
\label{tab:placement_delta}
\end{table}

\section{Conclusions}
Modern industrial-level floorplanning involves many placement constraints that govern the legal locations of many blocks. We have shown that popular floorplanning methods based on SA, however, typically fail to yield legal solutions in the presence of hard placement constraints. Our modified SA pipeline, CA-SA, introduces two novel mechanisms: constraints-fixing moves and \texttt{B*-trees} with anchored blocks. The former greatly increases the probability of finding solutions that satisfy boundary constraints while the later guarantees that pre-placement constraints are always satisfied. We introduced PARSAC, which is a parallel SA framework built on top of CA-SA. PARSAC can scale to many CPU cores and machines and can utilize the distributed compute resources to quickly find a rich Pareto front of {\bf legal} floorplans for complex floorplanning problems. 

Prior floorplanning work often failed to provide an accessible and properly documented code-base. This makes it hard, and sometimes impossible, to reproduce this prior work. 
We open-source the PARSAC library to allow others to reproduce our results. There is currently no easy-to-use open-source library for solving the floorplanning problem. PARSAC fills this gap, and introduces novel and effective mechanisms that allow users to move beyond toy floorplanning problems to problems with more realistic constraints reflective of modern SoC designs. 

\section{Acknowledgements}
\label{sec:acknowledgments}
We thank Miaomiao Ma (Intel), Pei Chun Ch'ng (Intel), Olena Zhu (Intel), and Fadi Aboud (Intel), for  bringing this problem to our attention and helping us understand the real-world constraints. 

\printbibliography

\end{document}